\documentclass[showpacs,superscriptaddress,twocolumn,amsmath,aps]{revtex4} 
\usepackage{graphicx}

\newcommand{\nl}{\nonumber \\}

\newcommand{\be}{\begin{equation}}
\newcommand{\ee}{\end{equation}}
\newcommand{\bea}{\begin{eqnarray}}
\newcommand{\eea}{\end{eqnarray}}

\newcommand{\Eq}[1]{Eq.\,(\ref{#1})}

\newcommand{\la}{\langle}
\newcommand{\ra}{\rangle}
\newcommand{\dg}{\dagger}

\newcommand{\mb}{\mbox}

\newcommand{\pone}{\small\mbox{$\frac{1}{2}$}}

\begin{document}
\draft

\topmargin=-40pt

\title{Quantum coherence control of solid-state charge qubit by means of
a suboptimal feedback algorithm}

\author{Jinshuang Jin}
\address{State Key Laboratory for Superlattices and Microstructures,
         Institute of Semiconductors, Chinese Academy of Sciences,
         P.O. Box 912, Beijing 100083, China}
\author{Xin-Qi Li}
\address{State Key Laboratory for Superlattices and Microstructures,
         Institute of Semiconductors, Chinese Academy of Sciences,
         P.O. Box 912, Beijing 100083, China}
\affiliation{Department of Chemistry, Hong Kong University of Science and
         Technology, Kowloon, Hong Kong}
\author{YiJing Yan}
\affiliation{Department of Chemistry, Hong Kong University of Science and
         Technology, Kowloon, Hong Kong}

\begin{abstract}
The quantum coherence control of a solid-state charge qubit is
studied by using a suboptimal continuous feedback algorithm within
the Bayesian feedback scheme. For the coherent Rabi oscillation, the
present algorithm suggests a simple bang-bang control protocol, in
which the control parameter is modulated between two values. For the
coherence protection of idle state, the present approach is
applicable to arbitrary states, including those lying on the equator
of the Bloch sphere which are out of control in the previous
Markovian feedback scheme.
\end{abstract}

\date{\today}
 \maketitle

{\it Introduction}.---
Being stimulated by the interest of solid-state quantum computation,
in recent years the measurement problem of solid-state qubit is
becoming an intensive research subject.
In particular, the setup of a charge qubit measured by a mesoscopic
transport device, for instance, the quantum point contact (QPC),
has received considerable attention \cite{Gur97,Ale97,Moz02,Gur03}.
Very recently, the mesoscopic QPC was experimentally demonstrated
as a charge-sensitive electro-meter \cite{Elz04}.

Strikingly, rather than the projective strong measurement,
the continuous weak measurement can
be employed to realize some marvellous tasks, such as quantum
state initialization, stabilization, and coherence protection, etc
\cite{Wis93,Ger04,Arm02}. The central idea is the
measurement-result-based feedback control, which falls typically
into two main categories, say, the Markovian scheme
\cite{Wis93} and the Bayesian scheme \cite{Bel92,Doh99}.
In quantum optics, the study of quantum feedback control has been going on
for more than a decade. However, in solid states it is a relatively new
subject \cite{Rus02,Kor05,Hop03,Rus95}.
In this work, based on a suboptimal feedback algorithm \cite{Doh01},
we study the problem of quantum coherence protection of a solid-state charge
qubit which is subject to the continuous weak measurement by a mesoscopic QPC.

{\it Formalism}.---
To be specific, the charge qubit is assumed to be two coupled
quantum dots (CQD's), occupied by a single electron \cite{Gur97,Kor99,Goa01},
whose Hamiltonian reads
$H_{qb}=\varepsilon(c_2^\dag c_2 - c_1^\dag c_1) + \Omega
(c_1^\dag c_2 + c_2^\dag c_1)$.
Here $c^{\dg}_{i}$ ($c_{i}$) is the electron
creation (annihilation) operator in the $i_{\rm th}$ dot,
$\varepsilon$ and $\Omega$ are, respectively,
the energy offset of the two local dot-states and their coupling strength.
At zero-temperature and in the limit of large bias voltage across the QPC,
the measurement back-action on the qubit is described by the following
master equation ($\hbar=1$) \cite{Goa01}
\begin{equation}\label{uncon}
\dot{\rho}(t)=-i[H_{qb}, \rho (t)] +
\mathcal{D}[\mathcal{T}+\mathcal{X} n_1] \rho(t) \;,
\end{equation}
where $n_1=c_1^\dag c_1$ is the occupation-number operator of the
first dot, the parameters $\mathcal{T}$ and $\mathcal{X}$ are given
by $|\mathcal{T}|^2=2\pi |T|^2g_lg_reV\equiv D_2$, and
$|\mathcal{T}+\mathcal{X}|^2=2\pi |T+\chi|^2g_lg_reV\equiv D_1$.
Here, the tunneling amplitudes through QPC, $T$ and $\chi$, are
assumed to be independent of the reservoir states, and real for
simplicity; $g_l$ and $g_r$ are the density of states for the left
and right reservoirs; $V=(\mu_L-\mu_R)/e$ is the bias voltage across
the QPC.
The superoperator in \Eq{uncon} is defined as $\mathcal{D}[r]\rho
= \mathcal{J}[r]\rho - \mathcal{A}[r]\rho$, where
$\mathcal{J}[r]\rho= r\rho r^\dag$, and
$\mathcal{A}[r]\rho=(r^\dag r\rho + \rho r^\dag r)/2$.
Intuitively, corresponding to the qubit electron in either the first
or the second dot, the QPC measurement current of
$I_1=eD_1$ or $I_2=eD_2$ is expected.
However, for continuous weak measurement, owing to the qubit electron jumps
between the two dots, the continuous output current $I(t)$ is a complicated
stochastic random quantity.
But, quite remarkably,
by using $I(t)$ one can design a proper feedback Hamiltonian $H_{fb}(t)$
to maintain or improve the quantum coherence of the qubit.
Conditioned on $I(t)$ and involving $H_{fb}(t)$,
\Eq{uncon} is unravelled as \cite{Goa01,Kor01,Sar04}
\begin{eqnarray}\label{Ito1}
d{\rho}_c(t)&=&-i[H_{qb},\rho_c]dt +
\sum_{j=x,y,z}\gamma_j\mathcal{D}[\sigma_j]\rho_c dt \nl
&& +\mathcal{D}[\small\mbox{$\frac{\mathcal{X}}{2}$}\sigma_z]\rho_c dt
+\mathcal{H}[\small\mbox{$\frac{\mathcal{X}}{2}$}\sigma_z]\rho_cdW(t) \nl
&&-i[H_{fb}(t),\rho_c]dt .
\end{eqnarray}
Compared to \Eq{uncon}, in addition to the feedback term (the fifth one),
we have included here the possible noise term (the second one)
due to external environment, which would spoil the quantum coherence
and needs to be eliminated by the feedback control strategy.
In \Eq{Ito1}, the superoperator ``$\mathcal{H}$" is defined by
$\mathcal{H}[r]\rho=r\rho+\rho r^{\dg}-\rho\; \mbox{Tr}[r\rho+\rho r^{\dg}]$.
$dW(t)$ is the Weiner increment, which is a Gaussian-white-noise stochastic
variable, having the property of $E[dW(t)]=0$
and $E[dW(t)dW(\tau)]=\delta(t-\tau)dt$, with $E[\cdots]$ the ensemble
average over a large number of stochastic realizations.
The Weiner increment $dW(t)$ is associated with the output current $I(t)$
in terms of the relation:
\bea\label{It2}
I(t)-\overline{I}=\Delta I
(\rho^c_{11}-\rho^c_{22})/2+\sqrt{S_0/2}\xi(t),
\eea
where $\xi(t)dt=dW(t)$, $\overline{I}=(I_1+I_2)/2$, $\Delta I=I_1-I_2$,
and $S_0=2e\overline{I}$.
$\rho^c_{ij}(t)=\la i|\rho_c(t)|j\ra$ is the density matrix element
over the qubit basis states.
Based on the output current $I(t)$, there are two typical schemes
to design the feedback Hamiltonian, i.e., the so-called Markovian
``$I(t)$-based" feedback \cite{Wis93},
and the Bayesian ``$\rho_c(t)$-based" feedback \cite{Bel92,Doh99}.
In the following, the second one will be implemented to protect the
qubit coherent evolution and its idle state, on the basis of a
suboptimal feedback algorithm  \cite{Doh01}.

\begin{figure}\label{Fig1}
\includegraphics*[scale=0.8,angle=0.]{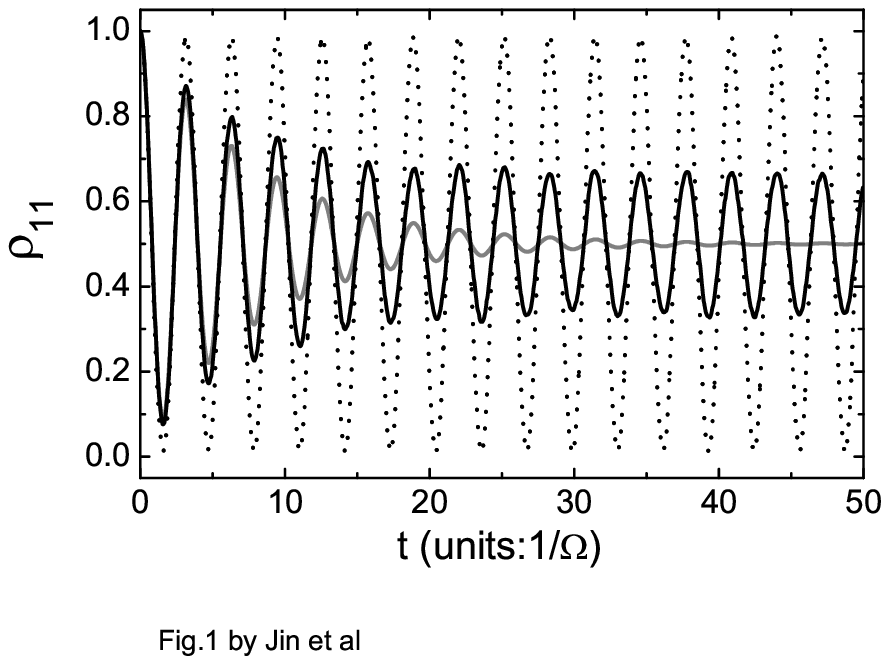}
\caption{The ensemble evolution of $\rho_{11}(t)=E[\rho^c_{11}(t)]$
for different feedback strengths $\mu$=0 (gray solid curve); 0.001
(black solid curve); 0.1 (dashed curve).}
\end{figure}

{\it Feedback Control of Rabi Oscillation}.---
For simplicity, consider the symmetric qubit ($\varepsilon=0$).
The ``desired" pure-state evolution to be protected is
\begin{equation}\label{target1}
|\psi_d(t)\rangle=\cos(\Omega t)|1\rangle -i\sin(\Omega
t)|2\rangle \;,
\end{equation}
where $|1\rangle$ and $|2\rangle$ are the dot states.
The basic idea of Bayesian feedback control is to carry out first
an {\it estimate} of the qubit state $\rho_c(t)$ based on the
noisy current $I(t)$, then compare it with the desired state $\psi_d(t)$
and design the feedback Hamiltonian using the calculated difference.
In Ref.\ \onlinecite{Rus02} the {\it estimated} quantity is the
{\it relative phase} between the superposed states ``$|1\ra$ and $|2\ra$", i.e.,
$\phi(t)=\arctan(2\mb{Im}\rho^c_{12}(t)/[\rho^c_{11}(t)-\rho^c_{22}(t)])$,
which is compared with $\phi_0=2\Omega t(\mod 2\pi)$
defined from the desired state.
Then, the phase ``error" $\Delta\phi(t)=\phi(t)-\phi_0$ is used to design
the feedback Hamiltonian, i.e., $H_{fb}(t)\propto \Delta\phi(t)$.
In certain sense, this scheme may be inconvenient to be implemented
in practice, since the feedback Hamiltonian is a complicated
stochastic function of time. In the following, we propose an
alternative scheme based on a suboptimal algorithm \cite{Doh01}.

For real-time feedback control, each feedback step
acts only for an infinitesimal time, ``$\Delta t$".
The suboptimal algorithm is based on the principle that
the state evolution in each infinitesimal time step will maximize
the fidelity of the estimated state with the desired (target) state.
The state evolution in the presence of feedback is governed by
\Eq{Ito1}. As far as the term related to the feedback Hamiltonian is
concerned, the final state $\rho_c(t+\Delta t)$ is given by \bea
\label{rhoc-dt}
  \rho_c(t+\Delta t)
  &=& \rho_c(t) - i[H_{fb},\rho_c(t)]\Delta t \nl
  &&-\pone [H_{fb},[H_{fb},\rho_c(t)]] (\Delta t)^2
   + \cdots ~.
\eea
\begin{figure}\label{Fig2}
\includegraphics*[scale=0.8,angle=0.]{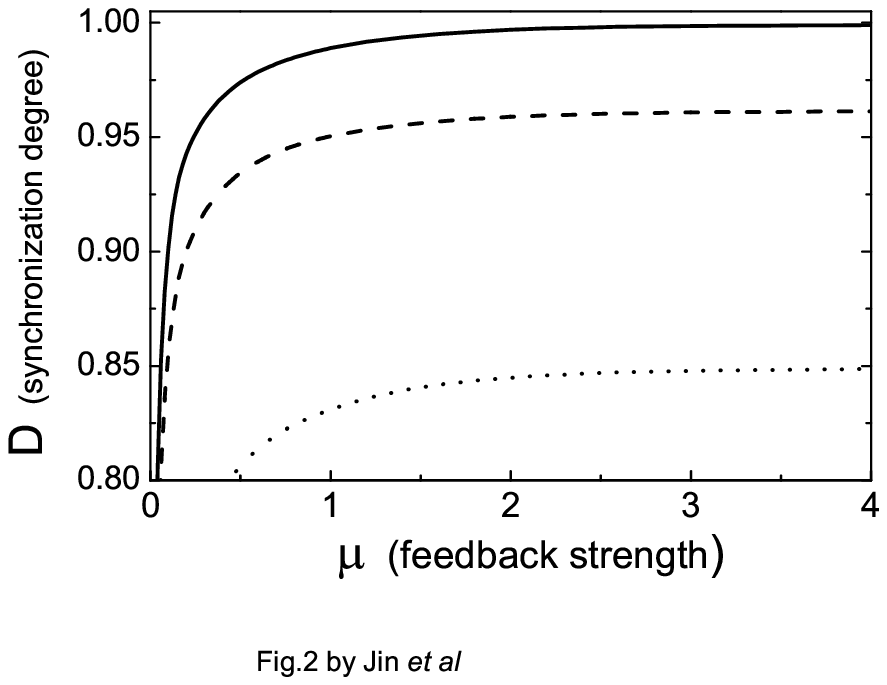}
\caption{ Synchronization degree $D$ as a function of feedback
strength $\mu$, for several magnitudes of dephasing due to
environment: $\gamma_z=$ 0 (solid curve); 0.05$\Gamma_d$ (dashed
curve); and 0.25$\Gamma_d$ (dash-dotted curve).}
\end{figure}
The fidelity of this state with the target state is defined by
\begin{eqnarray}\label{fidelity}
 \langle \psi_d| \rho_c(t+\Delta t)|\psi_{d} \rangle & = & \langle
 \psi_{d}|\rho_c(t)|\psi_{d} \rangle \nl
  & - & i\langle \psi_{d}|[H_{fb},\rho_c(t)]|\psi_{d} \rangle \Delta t \nl
  & - &  \pone \langle \psi_{d}|[H_{fb},[H_{fb},\rho_c(t)]]|\psi_{d}
 \rangle (\Delta t)^2 \nl
 &+& \cdots ~.
\end{eqnarray}
To optimize the fidelity, one should maximize the coefficient of $\Delta t$,
which is the dominant term.
Similar to other control theories,
the maximization must be subject to certain constraints,
such as the restriction on the maximum eigenvalue of $H_{fb}$,
the sum of the norms of the eigenvalues,
or the sum of the squares of the eigenvalues, etc. Physically, these
constraints stem from the limitation of the feedback strength
or finite Hamiltonian resources.
Here we adopt the last type of constraint, namely,
\begin{equation}\label{Hfb}
 \mbox{Tr}[H^2_{fb}] =  \sum_n \lambda^2_n(H_{fb}) \leq \mu .
\end{equation}
Under this constraint, the feedback Hamiltonian can be constructed
as \cite{Doh01}
\begin{equation}\label{Hfb-1}
  H_{fb} = i c [ |\psi_{d}(t) \rangle \langle \psi_{d}(t)|, \rho_c(t) ],
\end{equation}
where
 $   c = \sqrt{\frac{\mu}{2(a - b^2)}} $ ,
with
$  a  =  \langle \psi_{d}(t)| \rho^2_c (t)|\psi_{d}(t) \rangle $, and
$  b  =  \langle \psi_{d}(t)| \rho_c(t) |\psi_{d} (t)\rangle $.
Substituting \Eq{target1} into \Eq{Hfb-1} yields
\begin{equation}
H_{fb}=F\sigma_x \;,
\end{equation}
with
\begin{eqnarray}
  F & = & \left\{ \begin{array}{r}
  \sqrt{\frac{\mu}{2}} \;\; , \;\;\;  \; \Delta\phi < 0  \\
  -\sqrt{\frac{\mu}{2}} \;\; , \;\;\;  \; \Delta\phi > 0
                   \end{array} \right. .
\end{eqnarray}
In this case, the suboptimal feedback control resembles a simple
{\it bang-bang} control, where the feedback parameter between two
values is determined by the phase error $\Delta \phi$ as defined
above.
However, for other control purpose such as the protection of an
arbitrary idle state which is to be studied soon, the suboptimal
feedback scheme does not fall into the category of the ``bang-bang''
control, where the feedback Hamiltonian is more complicated.

\begin{figure}\label{Fig3}
\includegraphics*[scale=0.38,angle=0.]{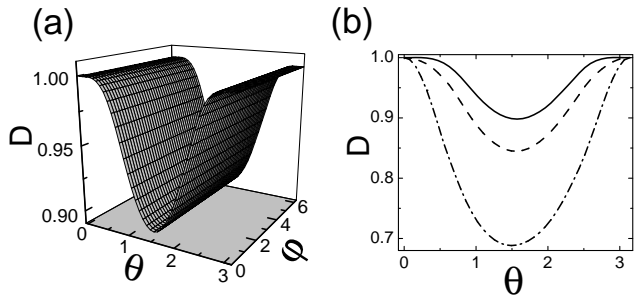}
\caption{(a) 3D-plot of the synchronization degree $D$ of the
feedback control for the measurement ; (b) 2D-plot of $D$ of the
feedback control for several magnitudes of dephasing due to
environment: $\gamma_z=$ 0 (solid curve); 0.05$\Gamma_d$ (dashed
curve); and 0.25$\Gamma_d$ (dash-dotted curve). The feedback
strength is $\mu=0.1$.}
\end{figure}

Now we present a Monte Carlo simulation for the feedback effect
based on the above algorithm.
In the simulation the coupling strength between the qubit and the detector
is characterized by the measurement-induced dephasing rate, which is assumed
$\Gamma_d=\mathcal{X}^2/2=0.25\Omega$.
In terms of the ensemble-averaged quantity $\rho_{11}(t)=E[\rho^c_{11}(t)]$,
Figure 1 shows that the simple feedback designed above can
really eliminate the back-action of the detector, where the ideal Rabi
oscillation is eventually recovered from the damped one
by increasing the feedback strength.
The result in Fig.\ 1 is carried out for the case of ideal detector
and in the absence of external noisy environment, i.e., by setting
$\gamma_j=0$.

To quantitatively characterize the effect of feedback,
the so-called synchronization degree is introduced as \cite{Rus02},
$D=2\langle {\rm Tr} \rho_c\rho_d\rangle-1$.
For instance, the complete synchronization is characterized by $D=1$,
indicating that the desired state is perfectly protected by the feedback.
In Fig.\ 2 we plot the synchronization degree versus the feedback
strength $\mu$, in the presence of external ``$\sigma_z$"-dephasing
with rates $\gamma_z=0$, $0.05\Gamma_d$, and $0.25\Gamma_d$, respectively.
In certain sense, the ``$\sigma_z$"-type error caused by the external
environment can be regarded as the consequence of a non-ideal detector
since the observable is also $\sigma_z$.
Figures 1 and 2 show that the fidelity of such feedback can be
arbitrarily close to 100\%, in the case of
using ideal detector and in the absence of external noises.
Nevertheless, it decreases in the case of a
non-ideal detector and/or in the presence of significant interaction
with external environment, as well as in the case of finite bandwidth
of the line which carries signal from the detector.
This limitation also existed in the scheme of Ref.\ \onlinecite{Rus02}.

{\it Protection of Idle State}.---
Now we turn to the subject of using the above suboptimal feedback
to protect the idle state of a qubit.
This is essential in information storage, while the Rabi oscillation discussed above
corresponds to information processing.
In general, an arbitrary pure state of qubit can be expressed as
\begin{equation}\label{state}
|\psi_T\rangle=\cos\frac{\theta}{2}|1\rangle+\sin\frac{\theta}{2}
e^{i\phi}|2\rangle \;.
\end{equation}
The goal of feedback is to make it immune against
the influence of environmental noise.

Following the same procedure used above, say, maximizing the ``$\Delta t$"
term in \Eq{fidelity}, and substituting \Eq{state} into \Eq{Hfb-1}, we obtain
\begin{equation}\label{Hfb2}
H_{fb} =\lambda[c_1\sigma_z+c_2\sigma_x+c_3\sigma_y] \;,
\end{equation}
with
\begin{eqnarray}\label{para}
\begin{array}{l}
\lambda=\sqrt{\mu/2(c^2_1+c^2_2+c^2_3)} \;,\nl
c_1 = 2\alpha\beta(\mb{Re}\rho^c_{12}\sin\phi+\mb{Im}\rho^c_{12}\cos\phi) \;, \nl
c_2=(\beta^2-\alpha^2)\mb{Im}\rho^c_{12}-\alpha\beta(\rho^c_{11}-\rho^c_{22})\sin\phi
\;,  \nl
c_3=\alpha\beta\cos\phi(\rho^c_{11}-\rho^c_{22})
-(\alpha^2-\beta^2)\mb{Re}\rho^c_{12} \;,
\end{array}
\end{eqnarray}
in which $\alpha=\cos\frac{\theta}{2}$, and $\beta=\sin\frac{\theta}{2}$.

So far the feedback Hamiltonian is constructed on the basis
of maximizing the coefficient of $\Delta t$ in \Eq{fidelity}.
However, this algorithm does not work if the target state is the eigenstate
of $\rho_c$, namely, $\rho_c|\psi_T\rangle=\lambda_T|\psi_T\rangle$.
For instance, such situation occurs when $\theta=0$ and $\pi$.
In this case, the dominant ``$\Delta t$"-term in \Eq{fidelity} vanishes,
and one is forced to maximize the coefficient of $(\Delta t)^2$.
Under the constraint on the Hamiltonian as described by \Eq{Hfb},
the explicit construction of the optimal $H_{fb}$ reads \cite{Doh01}
\begin{equation}\label{Hfb-dt2-1}
  H_{fb} = \sqrt{\mu/2} (|\upsilon_1\rangle \langle \psi_{T}| +
  |\psi_{T}\rangle \langle \upsilon_1|),
\end{equation}
where $|\upsilon_1\rangle$ is the eigenvector with the largest eigenvalue
of $\rho_c(t)$.
Notice that for the idle target state $|1\ra$ (or $|2\ra$), the
conditional evolution described by \Eq{Ito1} will not generate
quantum coherence between $|1\ra$ and $|2\ra$, because there is no
coherent driving source. Thus the states $|1\ra$ and $|2\ra$ must be
the eigenstates of $\rho_c(t)$, with eigenvalues $\rho^c_{11}(t)$
and $\rho^c_{22}(t)$, respectively.
From \Eq{Hfb-dt2-1}, the concrete form of feedback Hamiltonian is detailed as
\begin{eqnarray}\label{fb0}
  H_{fb} & = & \left\{ \begin{array}{r}  \sqrt{\frac{\mu}{2}}\sigma_x \;\; , \;\;\;
                                         \rho^c_{11} < \rho^c_{22}  \\
                                        0 \;\; , \;\;\; \rho^c_{11} > \rho^c_{22}
                        \end{array} \right. ,
\end{eqnarray}
for $ \theta=0$ ( $|\psi_T\rangle=|1\rangle$), and
\begin{eqnarray}\label{fbpi}
  H_{fb} & = & \left\{ \begin{array}{r}  0  \;\; , \;\;\;
                                         \rho^c_{11} < \rho^c_{22}  \\
                                        \sqrt{\frac{\mu}{2}}\sigma_x \;\; ,
                                        \;\;\; \rho^c_{11} > \rho^c_{22}
                        \end{array} \right. ,
\end{eqnarray}
for $ \theta=\pi$ ($|\psi_T\rangle=|2\rangle$).
Here we have used the property that
if $|\upsilon_1\rangle$ in \Eq{Hfb-dt2-1} is the target state,
we should set $H_{fb} = 0$ for that time step, since in this case
the fidelity will not increase under the feedback action.

In the following numerical simulations,
we first consider the problem of merely eliminating the back-action
of an ideal detector (i.e. $\gamma_j=0$), in the absence of
any other external influence of environment.
The result is shown in Fig.\ 3(a).
It is observed that the synchronization degree
is independent of the relative phase ``$\phi$" between the two superposed
components [c.f. \Eq{state}], but depends on ``$\theta$" symmetrically
around ``$\theta=\pi/2$" (the minimal point).
Since observing ``$\sigma_z$" will induce dephasing of the target state,
this leads to the smaller synchronization degree for the
more coherently superposed state.
For $\theta=0$ and $\pi$, the corresponding target states
are the eigenstates of the observable $\sigma_z$;
they are thus immune against the back-action of the detector,
leading to $D=1$ and $H_{fb}=0$ according to
the proposed feedback scheme.
On the other hand, for $\theta=\pi/2$, the corresponding target state is
$|\psi_T\rangle=\frac{1}{\sqrt{2}}(|1\rangle+e^{i\phi}|2\rangle)$,
which is the most coherently superposed state.
This state is destroyed most seriously by the detector's back-action,
whose synchronization degree is thus minimal.

\begin{figure}\label{Fig4}
\includegraphics*[scale=0.6,angle=0.]{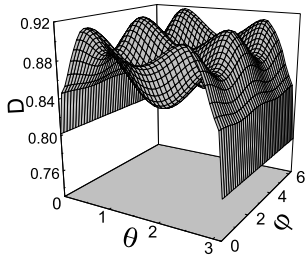}
\caption{3D-plot of the synchronization degree $D$ of the feedback
control
 for the $\sigma_x$-type error. The parameters are feedback strength $\mu=0.1$
 and noise strength $\gamma_x=0.05 \Gamma_d$.}
\end{figure}

Now we consider the influence of external environment.
In reality, a possible error is the $\sigma_z$-type,
which causes similar pure dephasing as the detector's back-action does.
As a consequence, the ``$(\theta,\phi$)"-dependence behavior of the
synchronization degree shown in Fig.\ 3 (b) is similar to Fig.\ 3(a),
but it decreases with the dephasing strength.
Another possible error is the $\sigma_x$-type, which causes relaxation
between the states $|1\ra$ and $|2\ra$.
The feedback result is shown by Fig.\ 4, where the synchronization degree
depends on both ``$\theta$" and ``$\phi$".
In particular, in contrast to the ``$\sigma_z$"-type error,
here for the states $|\psi_T\rangle=|1\rangle$ and $|2\rangle$
(corresponding to $\theta=0$ and $\pi$) the feedback efficiency is worst,
since they are the most seriously affected states by the
``$\sigma_x$"-type error.
Note also that the QPC tunneling amplitude related to $|1\ra$ is smaller
than that related to $|2\ra$, which leads to $D_{\theta=0}>D_{\theta=\pi}$.

{\it Conclusion}.---
To summarize, based on a suboptimal feedback algorithm we have
presented a study on quantum coherence control of a solid-state
charge qubit, for both its quantum evolution and idle state. For the
coherent Rabi oscillation, our study leads to a simple bang-bang
control feedback scheme, in terms of modulating a single control
parameter by only two values, which differs from the existing
Bayesian scheme \cite{Rus02} and may be easier to implement in some
cases. For the coherence protection of idle state, the present
approach is applicable to {\it arbitrary states}, including those
lying on the equator of the Bloch sphere which are {\it out of
control} in the previous  Markovian feedback scheme \cite{Wan01}.
In addition to eliminating the measurement back-action, the proposed
feedback scheme can also protect the qubit from environmental
influence {\it to some extent}. However, as many other approaches,
it cannot protect the qubit from {\it drastic influence} of external
environment.
This drawback may be partially overcome by developing other {\it
optimal} feedback algorithms, which is a promising but challenging
task in this field.

\vspace{3ex}
{\it Acknowledgments.}
Support from the National Natural Science Foundation of China
and the Research Grants Council of the Hong Kong Government
are gratefully acknowledged.





\end{document}